# 50 GHz Piezoelectric Acoustic Filter

**Omar Barrera* (Student Member, IEEE), Jack Kramer* (Student Member, IEEE), Lezli Matto, Vakhtang Chulukhadze (Student Member, IEEE), Sinwoo Cho (Student Member, IEEE), Michael Liao, Mark S. Goorsky, and Ruochen Lu (Senior Member, IEEE)**

The University of Texas at Austin, Austin, TX USA; University of California, Los Angeles, Log Angeles, CA, USA

CORRESPONDING AUTHOR: Jack Kramer (e-mail: kramerj99@utexas.edu).

*These authors contributed equally.

This work was supported by DARPA Compact Front-End Filters at the ElEment-Level (COFFEE)

**ABSTRACT** This paper presents significant frequency scaling of acoustic filter technology to 50 GHz. This achievement is enabled by the periodically poled piezoelectric film $LiNbO_3$ multilayer stack, in which piezoelectric thin-films of alternating orientations are transferred in sequence, thereby allowing efficient exploitation of high-order modes with high quality factor and coupling coefficient in a thicker piezoelectric stack. The demonstrated filter is comprised of twelfth-order symmetric mode lateral-field-excited bulk acoustic wave resonators (XBARs), built on a 4-layer periodically poled piezoelectric 128º Y-cut lithium niobate ($LiNbO_3$) stack. The filter exhibits 3.3 dB insertion loss and a fractional bandwidth of 2.9%. The miniature design, with a footprint of 0.36 $mm^2$, makes it promising for future wireless front-end applications. These results represent the highest frequency acoustic filters reported to date, setting a new benchmark in piezoelectric filter technology. Upon further development, the platform could enable filters further into the FR2 range, essential for next-generation communication systems.

## I. INTRODUCTION

The rapid proliferation of smartphones created a ripple effect that led to a growing demand for larger data transfer speeds, and consequently, the need for higher operating frequencies [1]. A key element that has contributed to the success of the smartphone is the introduction of miniature acoustic filters [2], [3]. In an acoustic device, electrical signals are converted to mechanical vibrations, which propagate with relatively low attenuation [4], [5]. Additionally, acoustic wavelengths are several orders of magnitude smaller than those of electromagnetic (EM) waves, enabling efficient energy transfer and compact design [6]. These characteristics make acoustic wave devices ideal for meeting the stringent size and performance requirements of radio frequency (RF) front end for modern mobile communication systems (Fig. 1). With the demand for data transfer rates projected to keep rising [7], acoustic devices are expected to remain the backbone of RF mobile filtering [8]. Two types of acoustic resonator technologies are often employed in acoustic filters, namely surface acoustic waves (SAW) and bulk acoustic waves (BAW) resonators. The latter is the preferred method for higher frequency applications, since BAW resonators maintain better electromechanical coupling ($k^2$) and quality factors ($Q$), required for high-performance filters [9]. Frequency scaling in BAW devices is implemented by reducing the thickness of the piezoelectric layer. While straightforward in theory, practical limitations exist due to fabrication constraints. Moreover, ultra-thin film devices are difficult to implement in 50Ω systems, and high acoustic damping results in reduced device performance [10]. As a result, commercially available devices based on aluminum

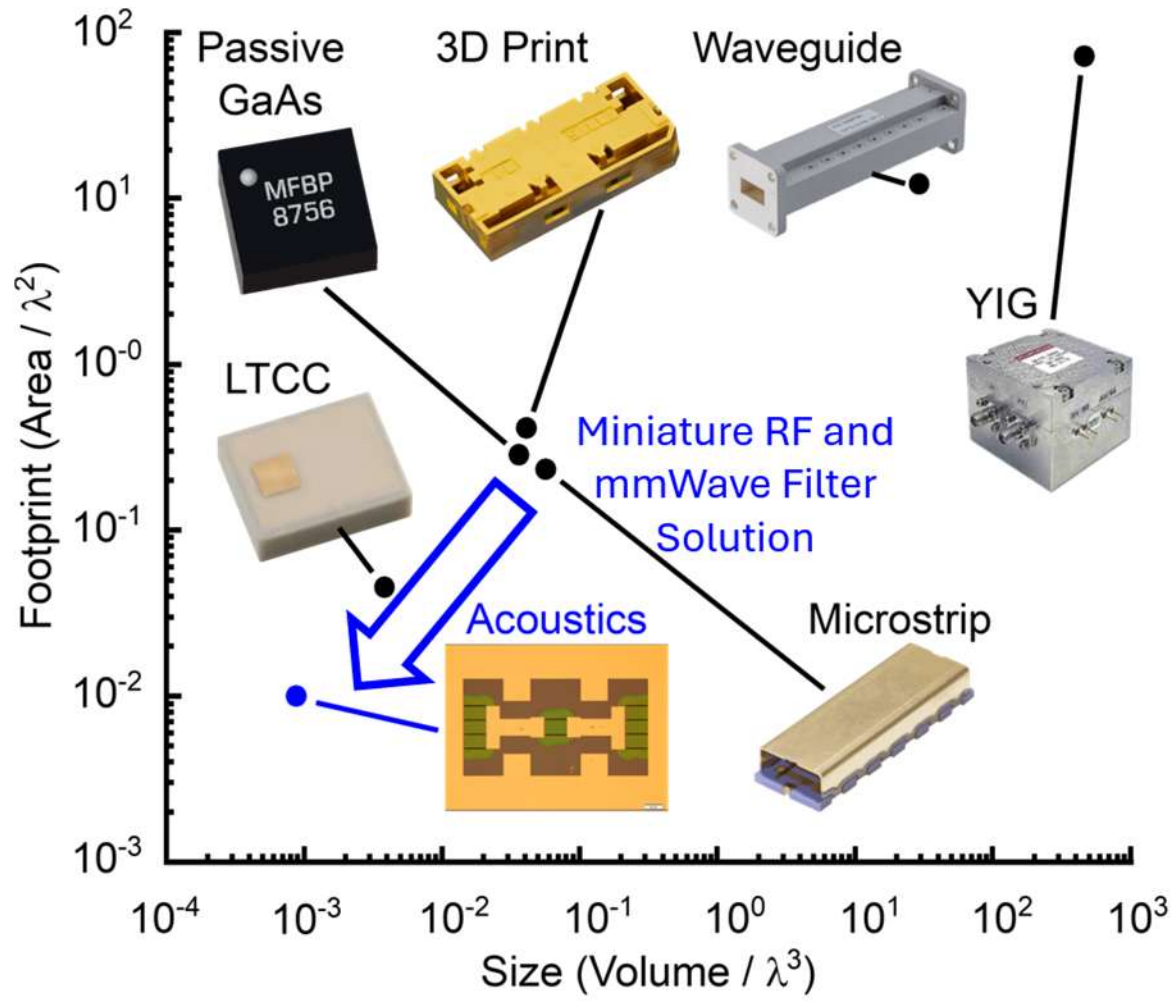

Fig. 1 Dimension comparison between different high-frequency filter technologies. The frequency ranges of each technology in this chart are: passive GaAs up to 37 GHz, 3D printed up to 52 GHz, waveguide to above 100 GHz, microstrip to 10 GHz, LTCC to 27 GHz, and YIG filters from 18-40 GHz. Acoustics were typically for sub-10 GHz until recent demonstrations.

Table I State-of-the-Art Acoustic Filters above 10 GHz

| Reference | $f_c$ (GHz) | *IL* (dB) | *FBW* (%) | Rejection (dB)* |
|---|---|---|---|---|
| [29] | 9.96 | 0.76 | 5.7 | 3.8 |
| [28] | 17.4 | 3.3 | 3.4 | 16.6 |
| [20] | 19.0 | 8.0 | 2.4 | 13.0 |
| [19] | 23.5 | 2.4 | 18.2 | 13.0 |
| [25] | 23.8 | 1.5 | 19.4 | 12.1 |
| [21] | 38.7 | 5.6 | 17.6 | 15.8 |
| **This work** | **50.1** | **3.3** | **2.9** | **8.9** |

**Rejection is measured at the highest out-of-band point.*

nitride/scandium aluminum nitride (AlN/ScAlN) have remained limited at sub-6 GHz frequencies [11], [12].

Recently, transferred thin-film lithium niobate ($LiNbO_3$) has been proposed as a new BAW platform [13], [14], [15]. These laterally excited resonators, known commercially as lateral-field-excited bulk acoustic wave resonator (XBAR) [16], excite the first anti-symmetric (A1) mode and provide large $k^2$ [17], operating deep into the mmWave frequency range [18]. The platform has successfully demonstrated high-performance acoustic filters in single layer $LiNbO_3$ above 10 GHz [19], [20]. Ultra-thin-film devices have also been reported, though with lower performance, highlighting the limitations of using single-layer stacks for large frequency scaling [21], [22].

To overcome the limitations of the single-layer approach, multi-layer periodically poled piezoelectric film (P3F) devices have been demonstrated in $LiNbO_3$ [23], [24], as well as on AlN/ScAlN-based devices [25], [26], [27], [28], [29]. This approach enables larger $k^2{\cdot}Q$ metrics at high frequencies, along with the added benefits of thicker film's capacitance density for 50Ω impedance matching. Recently, three-layered P3F $LiNbO_3$ has been shown to exhibit strong resonator figures-of-merits to above 100 GHz [24]. However, while high figure-of merit-resonators are required by filters, shifting from stand-alone resonators to full filters poses a significant hurdle at these frequencies. One of the significant challenges is the requirement to controllably shift the resonance frequency of resonators in the filter. In single layer thin films, this value is of the order of single nanometers. This is prohibitively difficult to implement practically. But by further increasing the piezoelectric layer count, the fabrication tolerance can be relaxed to physically realizable values. Such a platform provides the opportunity to shift high layer-count P3F $LiNbO_3$ from a resonator testbed to full compact filters operating in the mmWave.

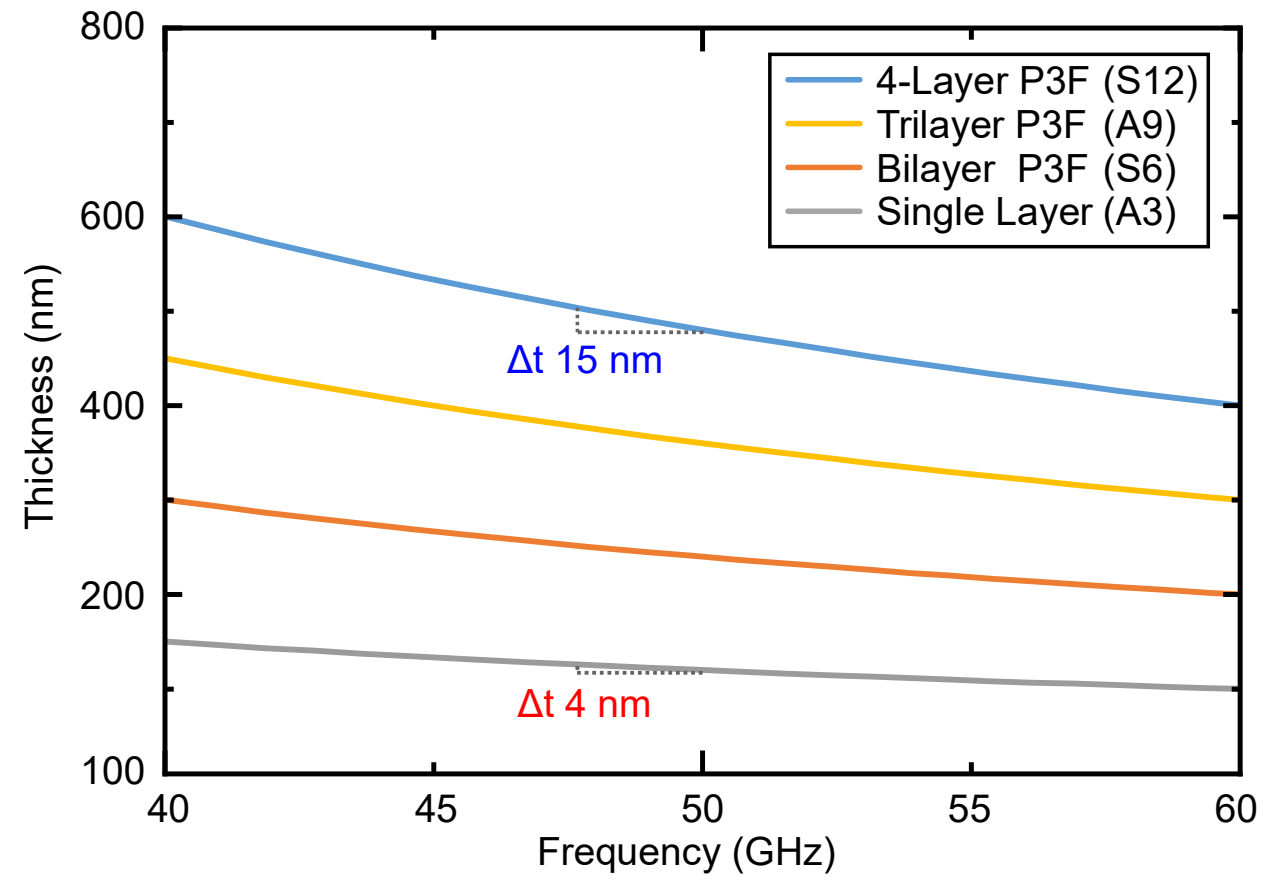

Fig. 2 Dispersion of A3 in $LiNbO_3$ XBARs and higher-order overtones in $LiNbO_3$ at 40-60 GHz, highlighting the relaxed requirement of fine trimming to realize the frequency shift for shunt and series resonators in thicker stacks, which is supported by multi-layer P3F $LiNbO_3$ for high FoM resonators.

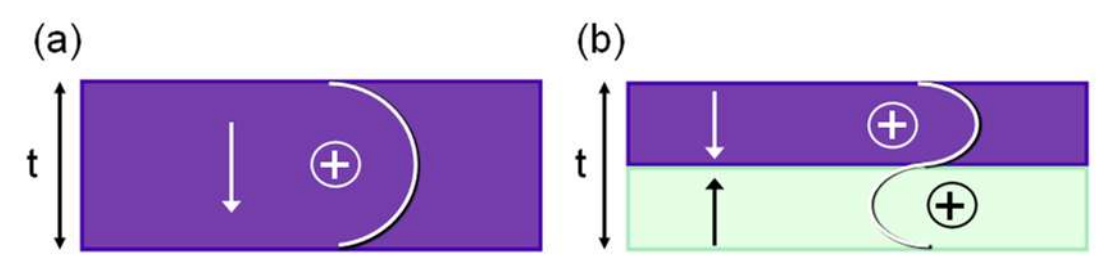

Fig. 3 Working principle of P3F stacks (a) single layer with thickness *t* operating at a frequency *f* and (b) bi-layer with thickness *t* and individual layers in opposite orientations. This mode operates at a frequency *2f*, with the same coupling coefficient as the fundamental in a single layer.

In this article, we report a 4-layer P3F filter at 50 GHz, significantly exceeding the current state of the art (Table I). The P3F filter's resonators utilize the 12th symmetric mode (S12) and exhibit $Q_s$ of 22 and 52, $k^2$ of 2.56% and 4.12% for the shunt and series resonators, respectively. The fabricated filter is centered at 50.1 GHz, with a low insertion loss (*IL*) of 3.3 dB and a 3-dB fractional bandwidth (*FBW*) of 2.9%. These results represent the highest frequency reported in the field to date, demonstrating piezoelectric filters deep into the FR2 bands for the first time [30].

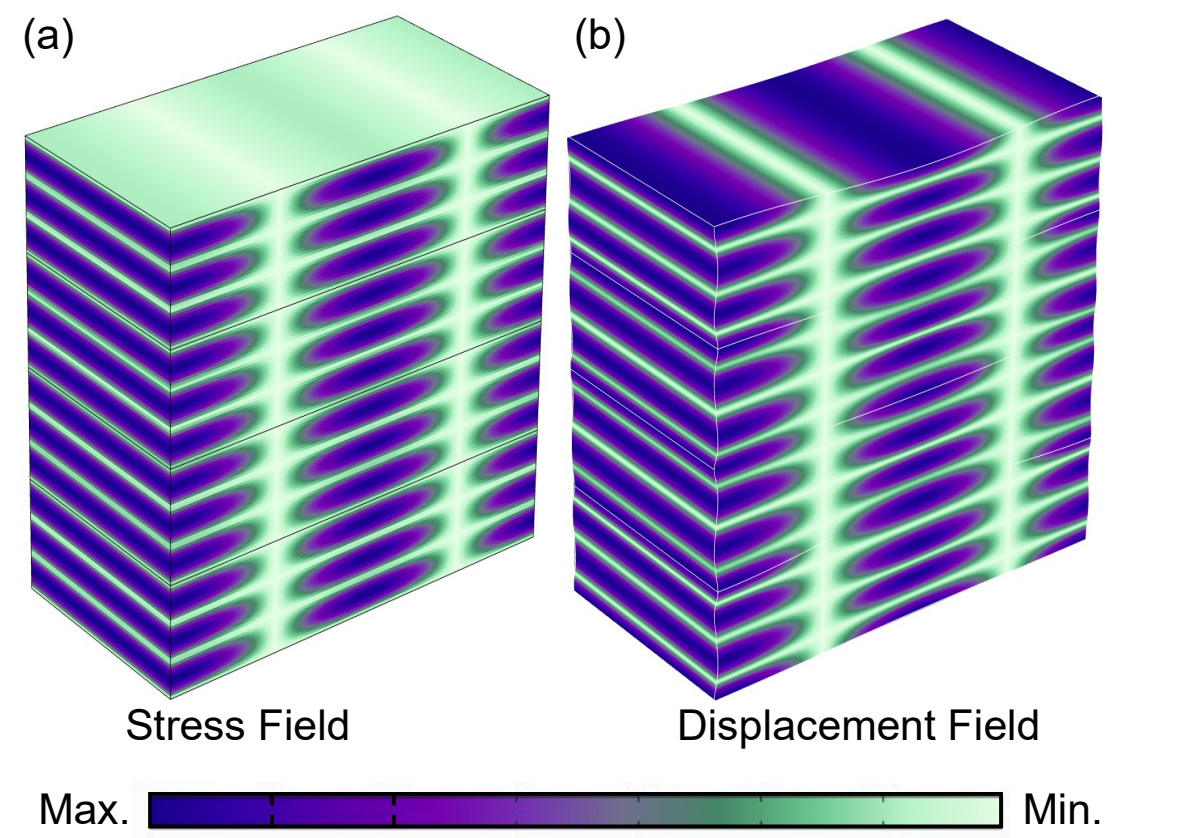

Fig. 4 FEA simulated (a) von Mises stress distribution and (b) displacement field of S12 mode in 4-layer P3F $LiNbO_3$.

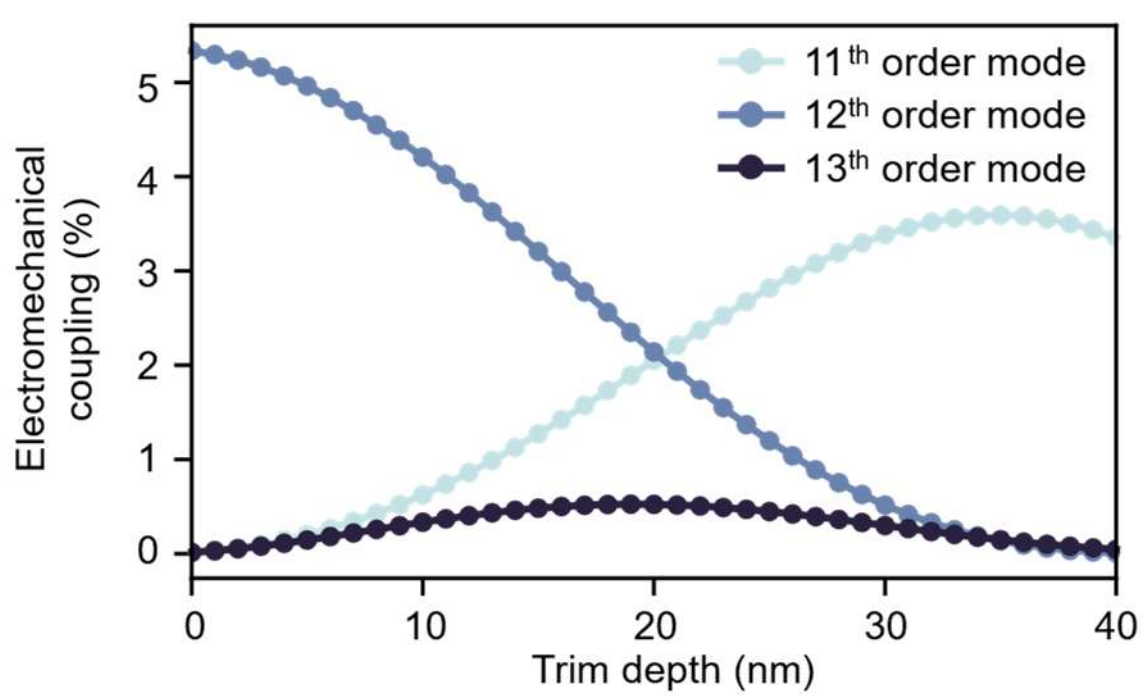

Fig. 5 Theoretical electromechanical coupling of an ideal 4-layer configuration for the targeted S12 mode and adjacent modes versus top layer trim depths.

## II. ANALYSIS AND DESIGN

The design of a mmWave acoustic filter at 50 GHz requires an appropriate piezoelectric material stack to support the desired acoustic resonances. A single-layer film working at the first symmetric (A1) mode would need to be approximately 40 nm thick, placing it in the lossy ultra-thin film regime. To overcome this task, utilizing a device employing its higher-order harmonics in a thicker stack presents a more practical solution. The calculated frequency response vs thickness for the proposed solution is displayed in Fig. 2 for a number of stacks. The frequency response is approximated by [31]:

$$f = \sqrt{(v_\lambda/\lambda)^2 + (N_h v_h/2h)^2} \qquad (1)$$

where $\lambda$ is the lateral wavelength, $v_\lambda$ is the acoustic velocity in the lateral dimension, $h$ is the thickness, $v_h$ is the acoustic velocity in the thickness direction, and $N_h$ is the mode order. This is an approximation of the Lamb wave frequency in a thin plate [32], [33], and illustrates the effect of film thickness on resonant frequency.

Based on Eq. 1, a single-layer stack employing the A3 mode at 50 GHz could be fabricated using a film of 110 nm thick (Fig. 2), which is within the reach of current fabrication capabilities. However, fine control over the

Table II. Resonator Design Parameters

| Resonator | $LiNbO_3$ Thickness (nm) | $\lambda$ (µm) | $f_s$ (GHz) | *Co* (fF) | *Q* | $k^2$ (%) |
|---|---|---|---|---|---|---|
| Series | 427 | 8 | 49.6 | 37 | 80 | 4.8 |
| Shunt | 440 | 8 | 48.4 | 80 | 80 | 7.5 |

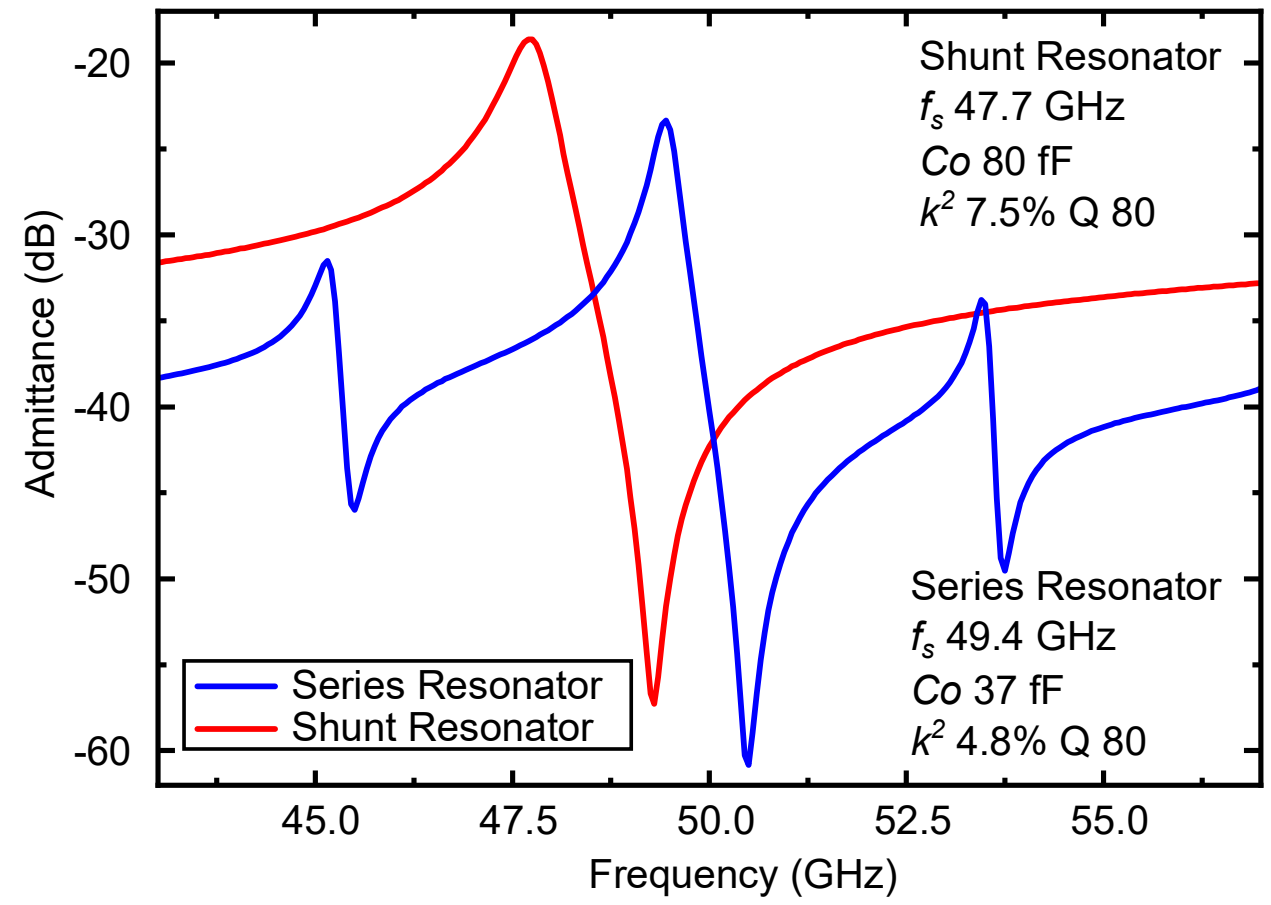

Fig. 6 Simulated shunt and series resonator admittance magnitude

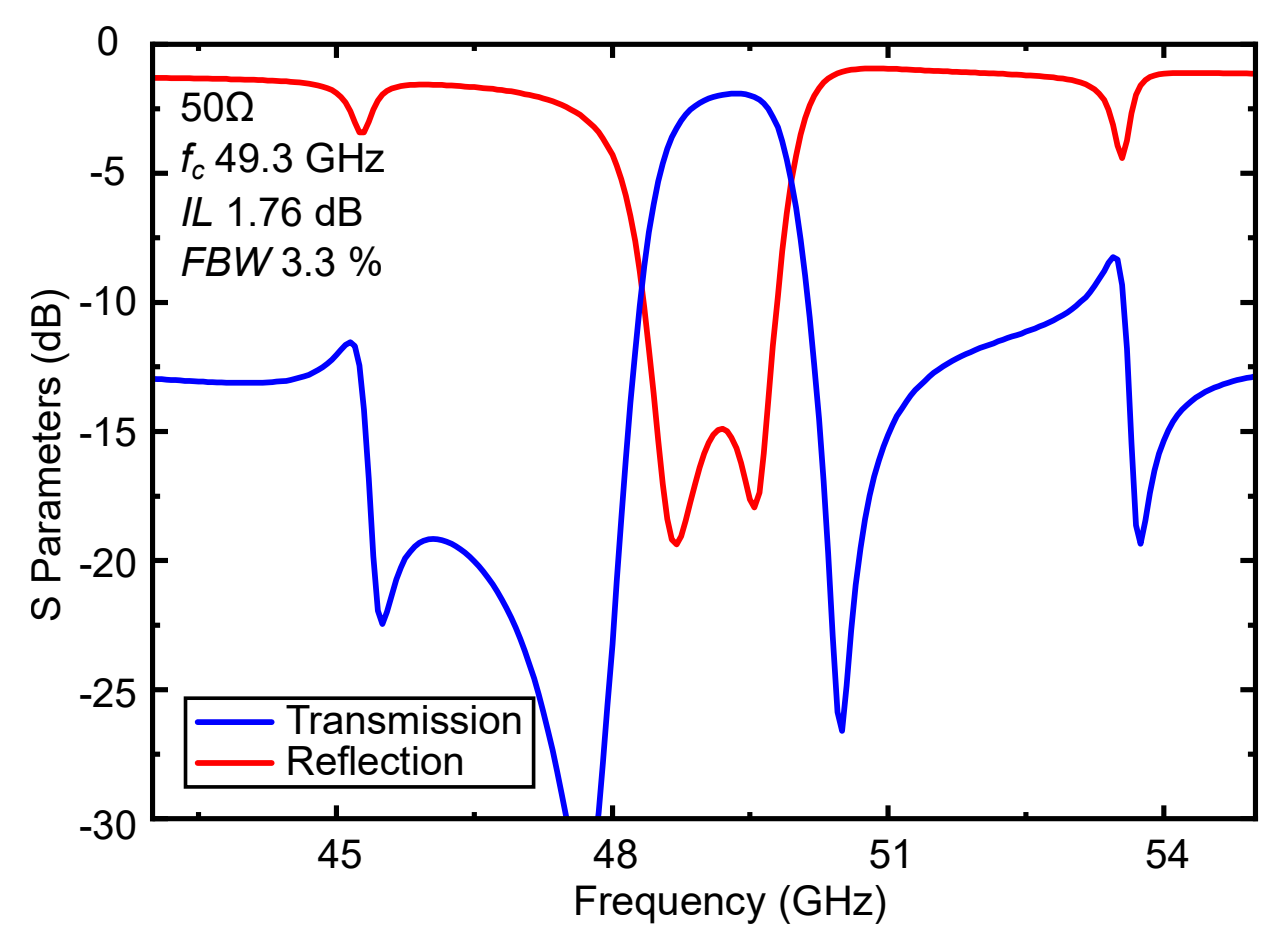

Fig. 7 Simulated 3rd-order ladder filter transmission and reflection.

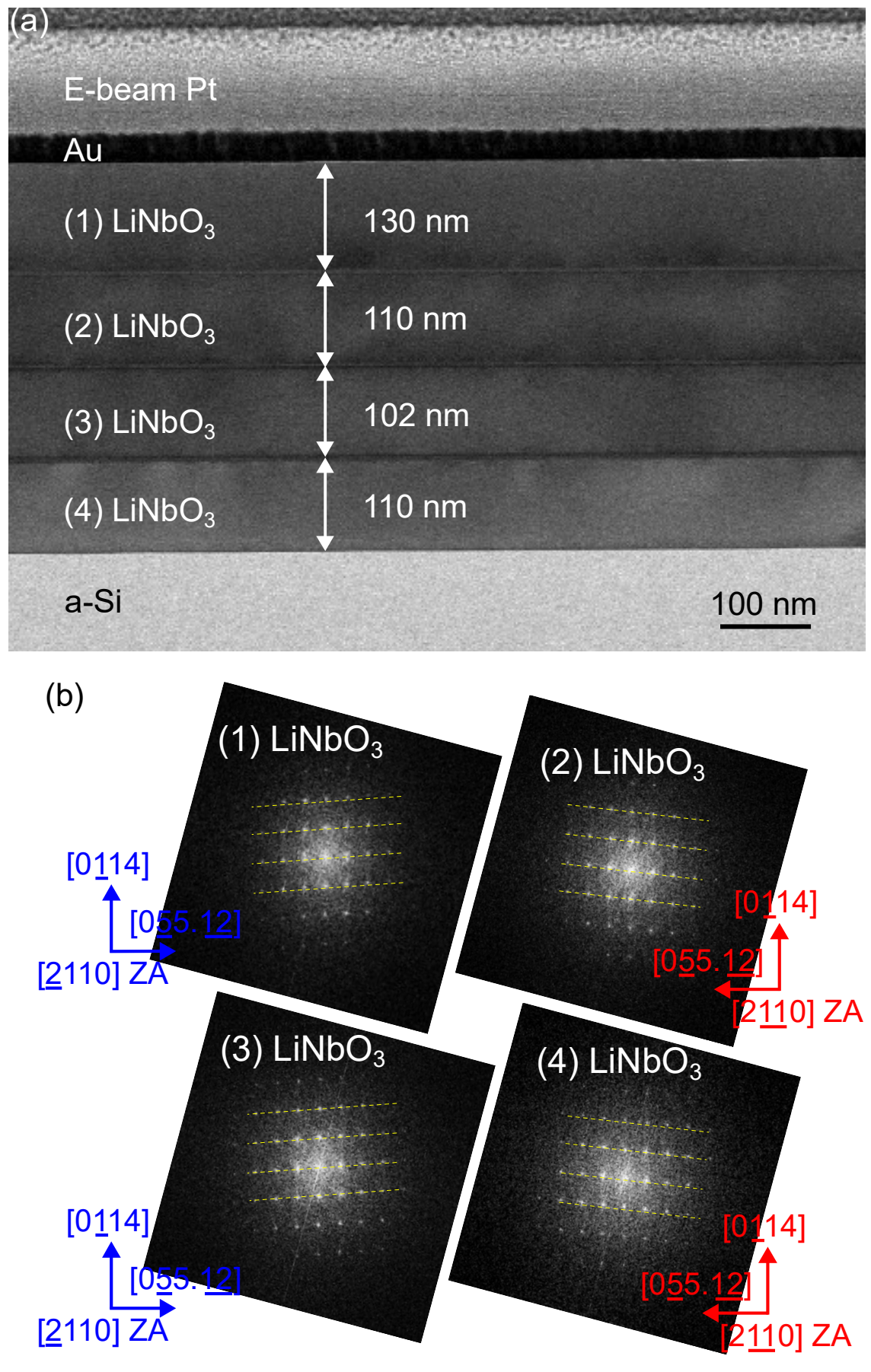

Fig. 8 (a) Cross-sectional BF-STEM image of the 4-Layer P3F $LiNbO_3$ and (b) FFT extraction of the reciprocal lattice of each layer.

frequency shift ($\Delta f$) for implementing shunt and series resonators poses a challenge: even a mere ±1 nm deviation can significantly degrade the filter performance, as the design requires a 4 nm thickness difference. Over etching can cause the series and shunt resonators to experience too large of a frequency shift, resulting in a notch in the filter passband. Under etching results in too small of a frequency shift, yielding a reduced filter bandwidth. One can further increase the mode order for more tolerance, but the $k^2$ of those resonators will be too low for filters.

As explained in the introduction, P3F layers with alternating piezoelectric orientations (Fig. 3) allow higher-order overtone operation without losing $k^2$. Consequently, a 4-layer P3F stack operating at the S12 mode can be fabricated with individual layers of 110 nm, resulting in a total thickness of 440 nm. This stack allows for a more manufacturable thickness difference of 15 nm needed to realize the required frequency shift, well within current ion-mill trim processing tolerances. The added volume of a thicker stack also reduces the fractional impact of acoustic surface losses, leading to higher resonator $Q$, and hence, improved filter $IL$. Moreover, the increased thickness offers larger capacitance density, reducing the footprint of the devices. As such, a 4-layer P3F stack is selected for this work.

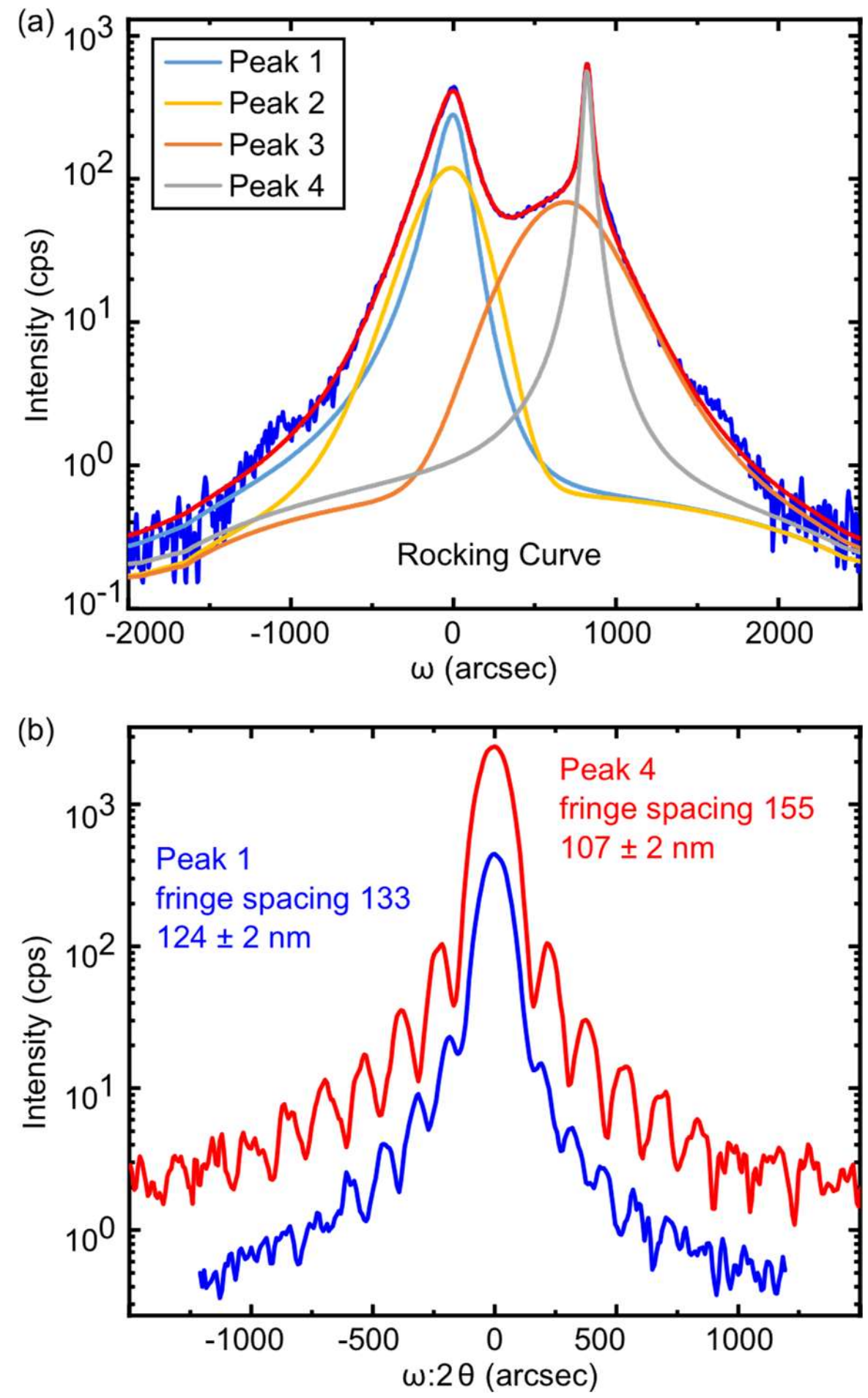

Fig. 9 (a) Rocking curve X-ray diffraction $\omega$ scan of the 4 layer lithium niobate platform. Peaks associated with each layer in the four layer stack are fitted to determine the individual peak widths. (b) X-ray diffraction $\omega$: $2\theta$ scan of 4 layer lithium niobate stack. Thickness fringes for peaks 1 and 4.

The design process begins by simulating the stack response using COMSOL finite element analysis (FEA). The resulting stress and displacement distributions for the 4-layer stack are shown in Fig. 4. The stress profile (Fig. 4a) shows a total of 12 alternating stress maxima along the thickness direction, indicative of the S12 mode. For the 4-layer P3F stack, S12 (three half-wavelengths per layer) was selected as the highest practical coupling mode near 50 GHz while maintaining manageable mode crowding; thickness matching minimizes spurious responses, and we therefore start with a slightly thicker top layer and trim within a controlled window to set Δf while preserving mode purity. In terms of displacement (Fig. 4b), particle motion is observed to alternate direction in the longitudinal axis, which is typical of shear wave propagation. It should be noted that the addition of electrodes to the FEA simulation causes a change to the modal profile that causes deviation in $k^2$ compared to the theoretical case.

The thickness difference for the resonators is selected from the FEA results in order to obtain the required frequency shift to realize the filter passband. FEA is required since equation (1) provides only a first order approximation of the resonant frequency, and does not incorporate important geometrical and material factors. The resonator design is largely based on the work reported in [18]. The design values are listed in Table II. For simulation purposes, the quality factor of the resonators are set to be 80, which is a representative value according to measurements of previously fabricated resonators [23]. A thickness difference of 13 nm is chosen here for the top layer $LiNbO_3$ trimming, slightly different from that calculated in Eq. 1, as the electrodes contribute to the resonance frequency. The frequency domain FEA is shown in Fig. 6. For the shunt resonator, S12 is the only tone in the spectrum. For the series resonator, the top layer will be trimmed down to increase the frequency of the series resonator. The reduced relative top layer thickness causes the adjacent 11th-order antisymmetric (A11) and 13th-order antisymmetric (A13) modes to be excited, as the stress field will no longer be fully cancelled due to nonuniform thickness [34], [35], but still with much lower coupling. A theoretical calculation of how the electromechanical coupling of the three modes changes versus trimming depth is shown in Fig. 5. This calculation makes use of the acoustic framework presented in [34], and assumes that all layers except the top layer have uniform 110 nm thicknesses. This theoretical analysis additionally shows the corresponding increased $k^2$ of the adjacent A11 and A13 modes.

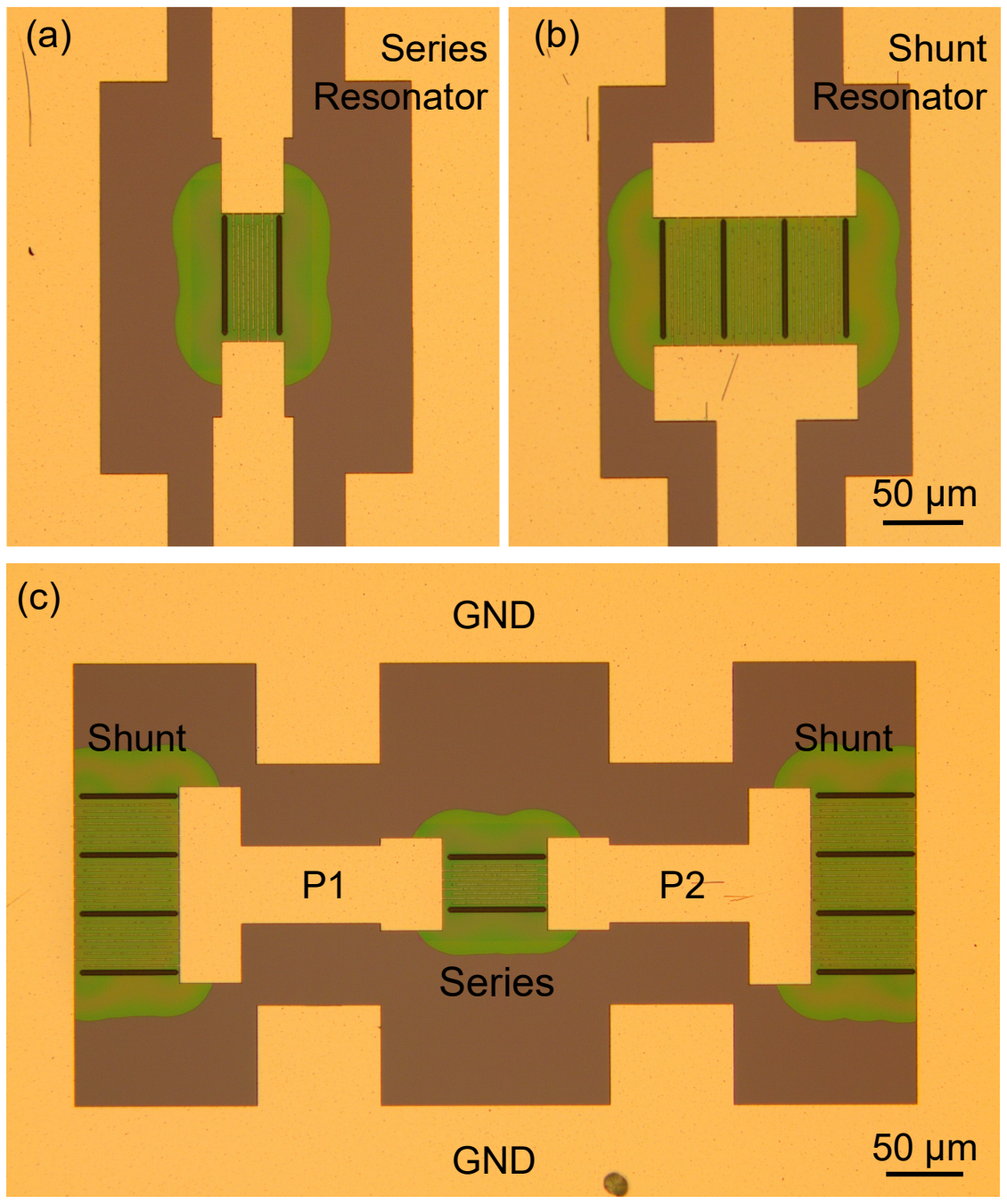

Fig. 10 Optical image of fabricated standalone (a) series resonator, (b) shunt resonator and (c) the full 50 GHz filter.

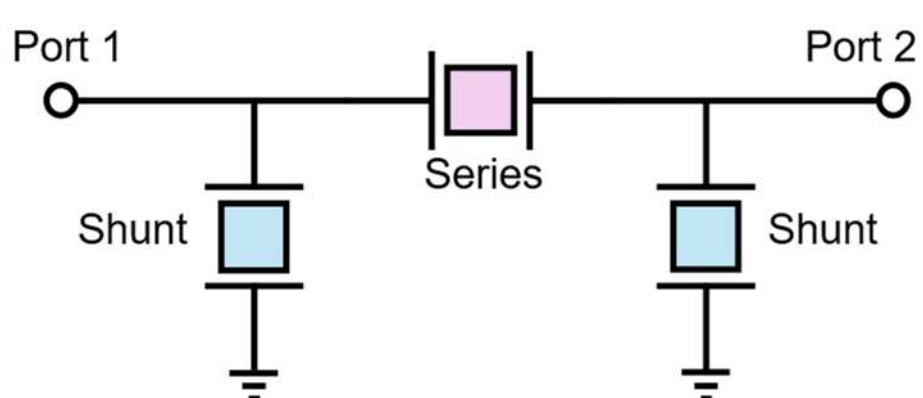

Fig. 11 Schematic representation of the 3rd order ladder filter shown in Fig. 10 (c).

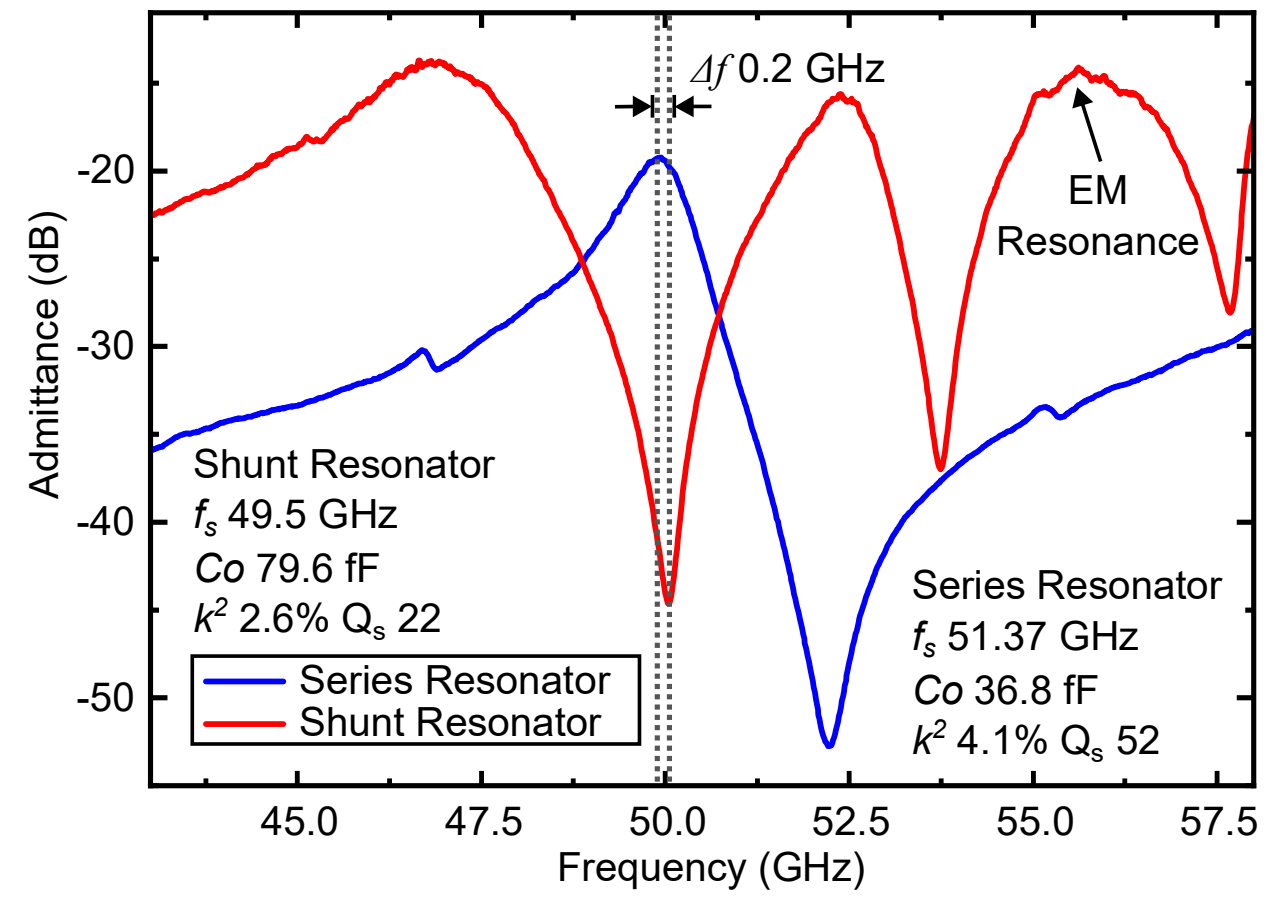

Fig. 12 Measured admittance magnitude of standalone resonators. An electromagnetic (EM) resonance is present in the shunt resonator due to the increased $C_0$, which forms an $LC$ with a parasitic series trace inductance.

The filter design is implemented by exporting the resonator performance from COMSOL and utilizing the modified Butterworth-Van Dyke (mBVD) model to fit the performance. The resulting parameters are then used to simulate a third-order ladder filter response in Keysight Advanced Design System. Here, we design the resonator parameters (e.g., $C_0$ and motional resistance, together with the targeted $f_s$ and $\Delta f$) such that the assembled ladder network realizes the desired transfer function under 50 Ω load terminations. The filter's impedance is tuned by adjusting the static capacitance of each resonator. It is worth noting that this impedance level is determined, to first order, solely by the electrode spacing. This provides robustness against process variations, since these dimensions are easily controlled during fabrication [36]. Notably, advanced filter-synthesis methods [37], [38], [39], [40] could be applied to refine the results further. The simulated filter response (Fig. 7) is centered at 49.3 GHz, exhibiting a low *IL* of 1.7 dB and a 3-dB *FBW* of 3.3%, indicating a promising mmWave acoustic filter. The simulated out-of-band rejection is around 12 dB, except for the tones introduced by A11 and A13 in the series resonator, as discussed above. These modes can be eventually mitigated by starting with a film with slightly thicker top layers, but we will start with a uniform thickness P3F stack for the prototype purpose.

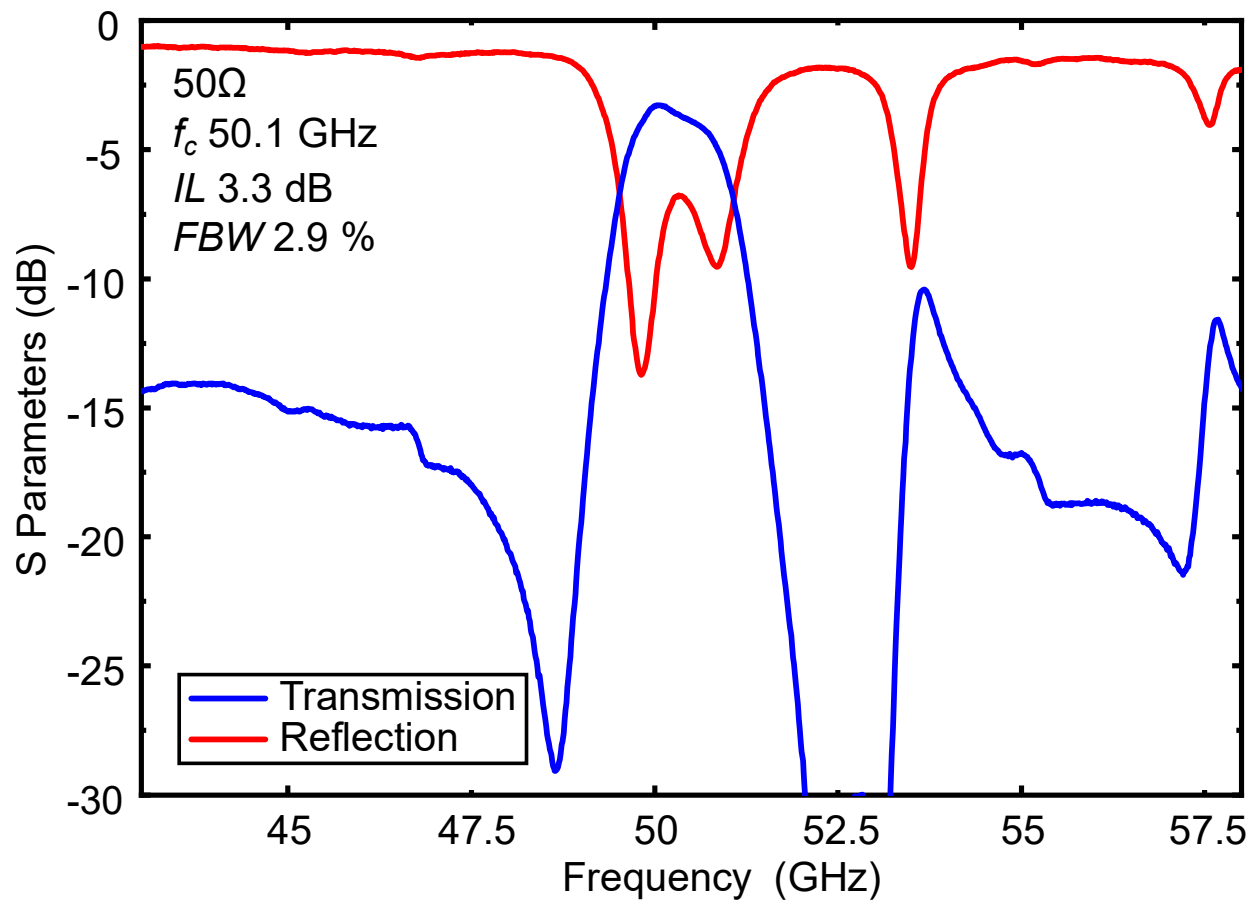

Fig. 13 Measured filter transmission and reflection response.

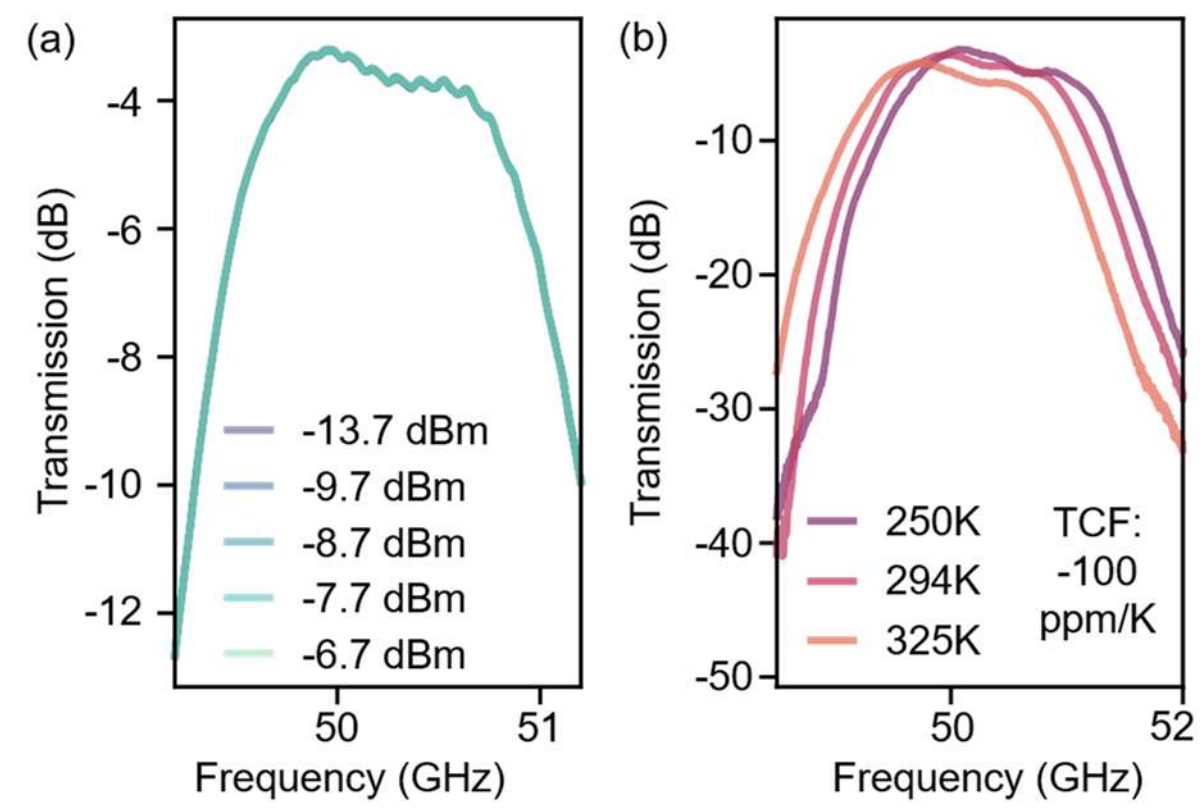

Fig. 14 (a) Measured filter response for different power levels, as seen at the power level at the probe tip. The lack of variation indicates a linear operation regime. Above this value, the vector network analyzer power becomes unlevel. (b) Select filter responses under variable temperature. The measured temperature coefficient of frequency (TCF) was measured to be -100 ppm/K. Variation in the passband insertion loss was caused by damage to the device probing pads.

## III. FABRICATION AND RESULTS

The stack is provided by NGK Insulators Ltd. It consists of a 4-layer P3F $LiNbO_3$ on 1 μm amorphous silicon (a-Si) layer, on top of a 500 μm sapphire substrate. Each adjacent thin layer of $LiNbO_3$ is rotated 180° about the material X axis and in-plane 180° with respect to adjacent layers to preferentially support the 4$^{th}$ order thickness shear mode [34]. Bright-field scanning transmission microscopy (BF–STEM) (Fig. 8(a)) shows the four distinct $LiNbO_3$ layers. The layers show only very small thickness differences, which could otherwise introduce additional modes in the frequency response of resonators [41], but are sufficient for the 50 GHz filter prototype. FFT extractions of the lattice for each layer are used to verify in-plane crystal orientation, showing alternating patterns for each subsequent layer. Along this [$2\underline{1}\underline{1}0$] zone axis of the $LiNbO_3$ layers, the in-plane lattice rotations are clearly visible and confirm the 180 ± 3° twist from layer to layer (Fig. 8 (b)).

High resolution X-ray diffraction characterization was performed with a Bruker-JV D1 X-ray diffractometer using triple-axis diffraction, which provides a non-destructive means to determine materials parameters of the layers. The triple-axis diffraction rocking curves of the symmetric (0114) reflection were used to quantify lattice tilt and mosaicity of the individual layers, and ω:2θ scans were used to measure layer thickness values. Peaks were separated along the rocking curve (Fig. 9(a)) scanning axis, and the peaks of each layer were deconvoluted by peak fitting. The peak width measured full width half maximum of 170”, 360”, 570” and 50” for layers 1, 2 and 3 and 4 respectively (using the layer identification from the TEM measurement in Fig. 9). Broader full width half maximum in the rocking curve corresponds to lattice tilt and mosaicity – these values were comparable to those from our prior publications. Fringe spacings from the ω:2θ scans (Fig. 9 (b)) were measured to determine thicknesses of two of the layers, which were found to be ~124 nm for peak 1, corresponding to layer 1 measured with STEM, and ~107 nm for peak 4, corresponding to layer 4 measured with STEM. The other two peaks did not exhibit enough intensity to resolve the thickness fringes.

Filter fabrication begins by defining local regions lithographically across a sample of 2.1 by 1.9 cm in dimensions. These regions are then selectively thinned by 13 nm using ion beam-assisted argon etching (ion-beam). The etched areas serve as a platform for the series resonators, with the thickness difference providing the required $\Delta f$. To achieve the trimming control required for this design, we implemented a multi-step ion beam etch. Each etch aims to remove only a 3-4 nm, at which point the sample is removed and the thickness change is measured using a Filmetrics F50-UV. Repeating this procedure provides control of the trimming depth to +/- 1 nm. This process has been demonstrated to maintain surface roughness following the etch [42]. Future studies will focus on the process's uniformity and trimming accuracy. Following, the metal layer is defined for electrodes and interconnects using electron beam lithography (EBL). Aluminum (Al) is deposited in two stages by evaporation: first, a 350 nm layer is deposited for both electrodes and interconnects, followed by an additional 350 nm for interconnects only, to enhance measurement robustness. Next, etch windows are patterned with EBL and etched deep into the aSi layer using ion-milling. Finally, the resonators are suspended using a silicon selective etch via xenon difluoride. The optical images of the fabricated filter and resonators are shown in Fig. 10, with a schematic representation of the resonator configuration shown in Fig. 11.

The devices are measured using an Agilent E8631 vector network analyzer in air at a power level of −15 dBm. A short-open-load-through method was used to calibrate the response at the input of the devices. The resonator's admittance magnitude is displayed in Fig. 12, the frequency offset $\Delta f$ between resonators shows good agreement with the values predicted from the dispersion analysis, validating the selection of the P3F stack. Additional acoustic modes are observed in the frequency response and are a result of thickness mismatches between layers. This effect is undesirable as it introduces points of low out of band rejection. But this also is a byproduct of the nascent phase of this technology, which can be corrected through commercialization and improved fabrication processes [43]. Furthermore, a resonance of EM nature is observed in the shunt resonator, arising from parasitic inductances in the layout. The parasitic inductances form an $LC$ tank that introduces an additional electromagnetic resonance. This occurs at a lower frequency due to the larger static capacitance, $C_0$, required to achieve an impedance matched filter.

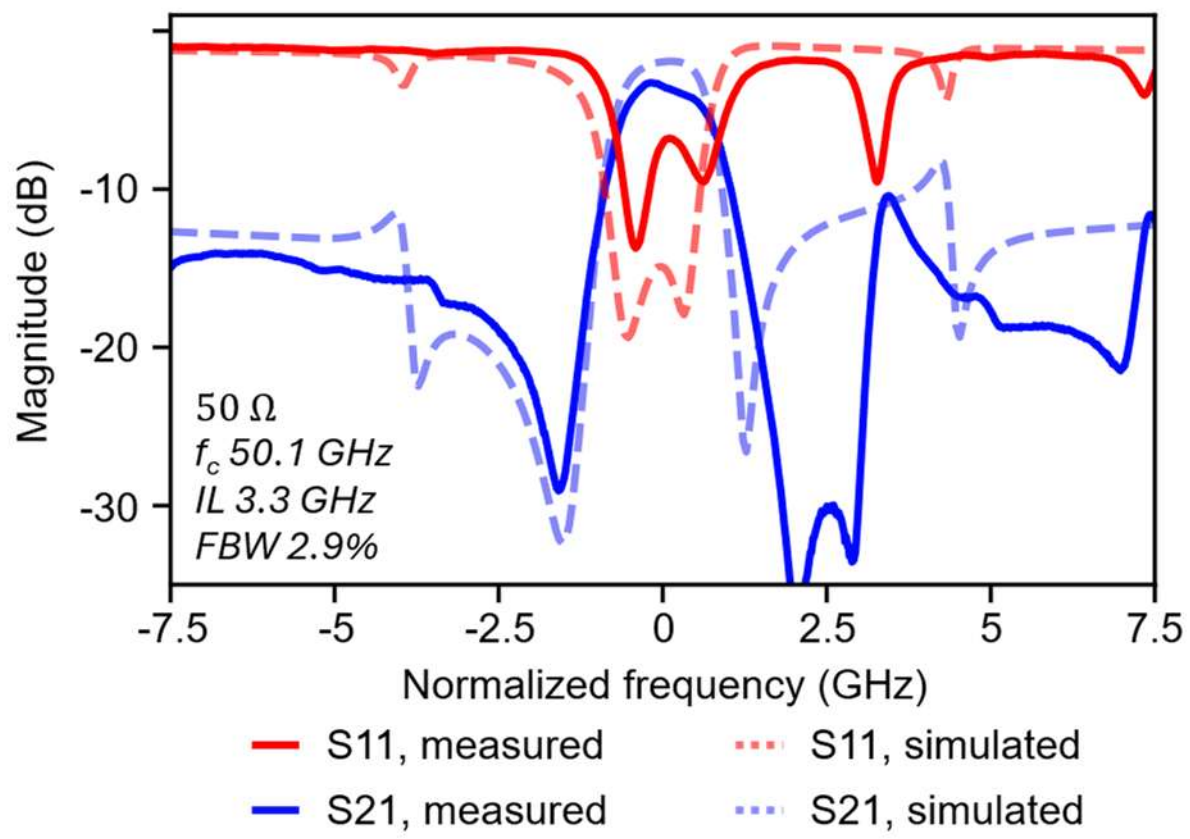

Fig. 15 Comparison of measured and simulated filter response after normalizing to the center frequency of the passband. Variation is attributed to the layer mismatch between the ideal (simulated) case and the practical stack.

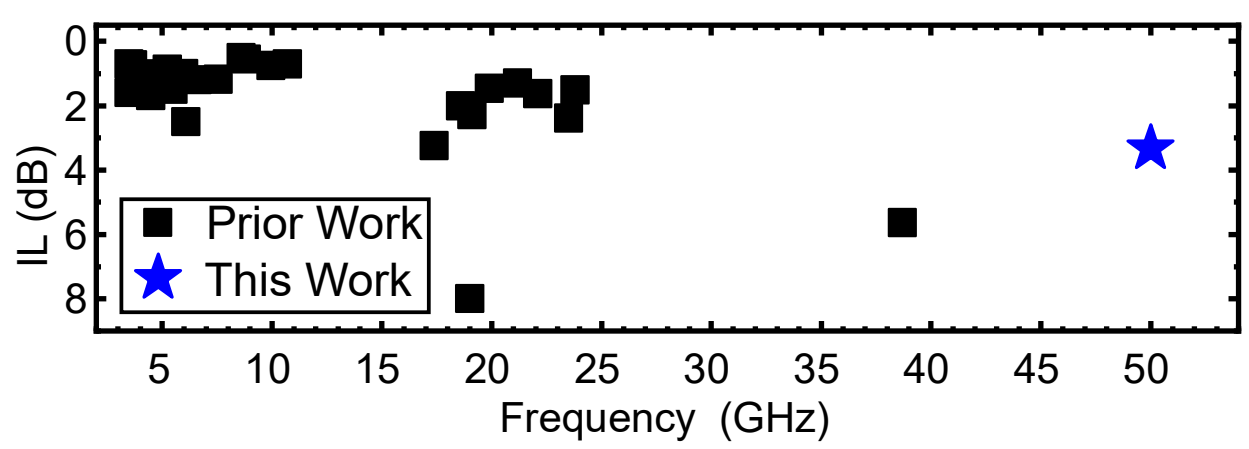

Fig. 16 Comparison of measured insertion loss (IL) of this filter to other reported acoustic filters.

The measured filter response, presented in Fig. 13, exhibits a passband centered at 50.1 GHz with an IL of 3.9 dB and a 3-dB FBW of 2.9%. These results are in agreement with the simulation response, demonstrating an effective utilization of an overtone in the P3F resonator stack to achieve a filter with a passband response in the upper FR2 frequency range. In comparison to other filter technologies at similar frequencies (Fig. 1) [44], [45, p. 3], [46], [47], [48], [49], this device features a bandpass response over a miniature footprint of 0.36 $mm^2$, and a volume of 0.19 $mm^3$, as predicted by the intrinsic short wavelength of the acoustic technology. The nonlinearity of the device has been explored. The device shows minimal performance difference from lower power levels up to -6.7 dBm as the incident power at the probe tip (Fig. 14). The current experimental setup does not support higher input power testing. Further exploration requires future works focusing on nonlinear characteristics, e.g., 1-dB compression point (P1dB), third-order input intercept point ($IIP_3$), or harmonic distortion, upon future setup availability. The temperature coefficient of frequency (TCF) was measured by performing a temperature sweep from 250K to 325K with 5 temperature points, at which point the probes caused excessive damage to probing pads. To extract the TCF, the 3 dB bandwidth was used to determine the filter center frequency, then a linear regression was performed to extract the slope versus temperature. For this configuration, the filter TCF was measured to be -104 ppm/K. This number could be improved through introducing temperature compensating materials such as silicon dioxide. Figure 15 compares the simulated ladder-filter response (center frequency 49.3 GHz, IL = 1.76 dB, 3-dB FBW = 3.3%) directly with the measured response (center frequency 50.1 GHz, IL = 3.3 dB, 3-dB FBW = 2.9%), after normalizing the center frequency. The overall passband shape and bandwidth show good agreement, with a modest ~0.8 GHz frequency shift and higher measured insertion loss. We attribute the residual differences primarily to process/trim tolerances that perturb fs and Δf, and to additional loss/parasitics from interconnects and the probing environment that are not fully captured in the simplified MBVD-based simulation. The slight difference in layer thicknesses of each layer leads to changes of the spurious mode position and amplitude. While the present demonstration is limited by the achievable layer-thickness control of the transferred-film stack, improved manufacturing uniformity and repeatability are expected to further reduce spurious responses and enable broader optimization across adjacent higher-order modes.

This paper marks the first demonstration of piezoelectric acoustic filter technology at 50 GHz, compared to the state of the art plotted in Fig. 16 and Table 1. This initial work highlights potential challenges, but also opportunities for mmWave piezoelectric filters. Upon development, especially in lowering IL and providing higher out of band rejection with closely integrated EM structures, such acoustic elements could provide new options for future mmWave front-end microsystems.

## IV. CONCLUSIONS

We have presented the analysis, design, and implementation of an acoustic filter centered at 50 GHz, marking a significant advancement in frequency scaling. This achievement is made possible by a four-layer P3F $LiNbO_3$ resonator stack that effectively harnesses the S12 overtone. The results demonstrated represent the highest operating frequency reported for acoustic filters to date, establishing a new benchmark in piezoelectric filter technology. This demonstration opens the doors for improvement in millimeter-wave acoustic filter design. Through introducing additional materials, such as silicon dioxide, the temperature response of the filters could be improved [50]. Localized focus ion beam trimming could provide better frequency control and reliability. Because the local top-layer trimming used to set Δf also introduces a slight layer-mismatch condition that can modify the effective $k^2$ [34], [51], future designs can incorporate additional knobs, such as in-plane resonator and electrode rotation [52], to more independently tailor $k^2$ for wider-band filters. And commercialization of the P3F platform could similarly bring increased control of the piezoelectric layers and interfaces, mitigating spurious modes and improving filter insertion loss. To address the currently modest out-of-band rejection, electromagnetic and acoustic codesign becomes important. This could help to improve the overall filter response through adding microwave structures such as microstrip inductors [53], [54]. Future work could incorporate formal filter-synthesis approaches as a starting point, with practical constraints (e.g., achievable $k^2$, loss, and parasitics) included during subsequent optimization and layout implementation. With these improvements, compact piezoelectric filters could become prominent component for millimeter-wave communications.

## ACKNOWLEDGMENT

The authors thank the DARPA COFFEE program for funding support, and Dr. Ben Griffin, Dr. Todd Bauer, and Dr. Zachary Fishman for helpful discussions. The authors also thank NSF CAREER: Radio Frequency Piezoelectric Acoustic Microsystems for Efficient and Adaptive Front-End Signal Processing.

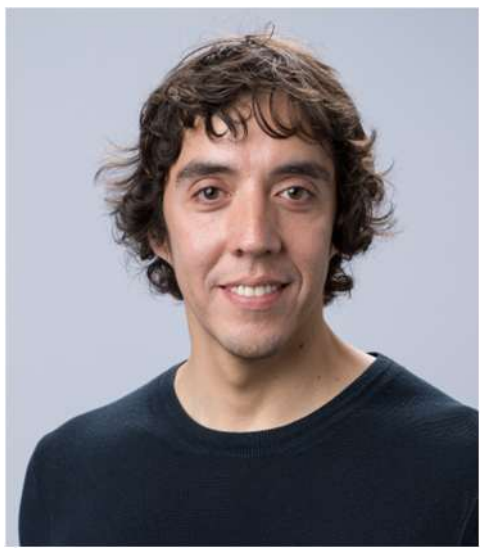

**Omar Barrera** (Student member, IEEE) Omar is a Ph.D. candidate at UT ECE. He graduated with a B.S. degree from UABC in Mexico and an M.Eng. from Dongguk University in Korea, both in Electrical Engineering. He has previous research experience in antennas and RF circuits. He also has industry experience in the automotive industry, where he worked as a manufacturing engineer. His current research interests involve microwave and mm-wave devices and circuits, and more in particular, the development and characterization of acoustic filters for new radio mobile applications. Omar was the recipient of an Outstanding Student Oral Presentation Award Finalist in IEEE MEMS 2024 and was awarded 2nd place in the Conference Paper Award in IEEE International Microwave Filter Workshop (IMFW 2024). He has also received the Cadence Diversity in Technology Scholarship in 2023.

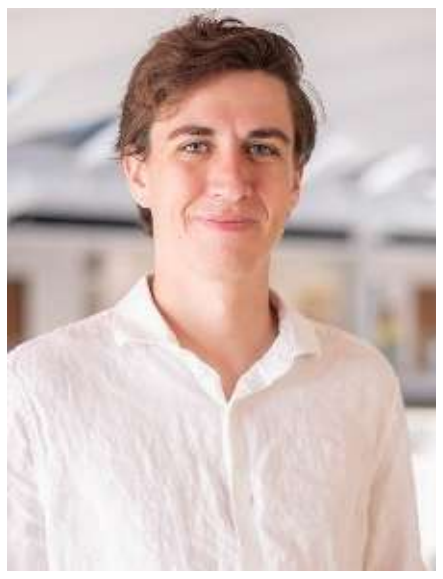

**Jack Kramer** (Student member, IEEE) received his B.S. from the University of New Mexico in 2021 and is currently pursuing his Ph.D. in electrical engineering at the University of Texas at Austin. His research has focused on millimeter-wave and sub-terahertz acoustic systems leveraging lithium niobate thin films. His interests include the further integration of these systems into hybrid optical and quantum systems. He received best student paper awards at IEEE International Frequency Control Symposium (IFCS) in 2023 and IEEE International Conference on Microwave Acoustics and Mechanics (IC-MAM) in 2022. He also received the 2024 Ben Streetman Junior Researcher Award, the 2025 University of Texas Continuing Fellowship, and the 2025 Texas Quantum Institute Fellowship.

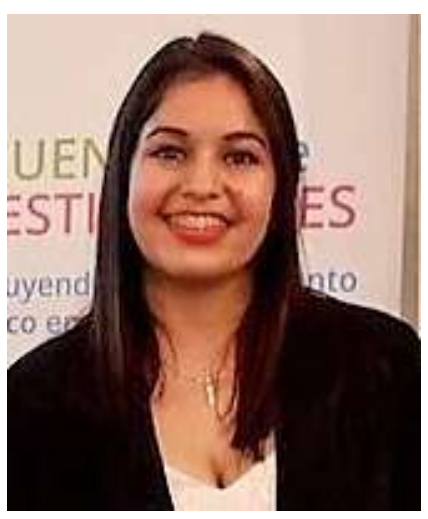

**Lezli Matto** received her B.S. in Materials Science and Engineering from the Polytechnic School at the National University of *Asunción* in Paraguay in 2017. She received her Master of Science in Engineering in Materials Science and Engineering from the University of Texas at Austin in 2021 funded by the Fulbright Scholarship. Lezli is currently a PhD candidate in the Materials Science and Engineering department at the University of California los Angeles (UCLA). Her research interests focus on wafer bonding, ion implantation, exfoliation and thin film layer transfer for heterogeneous integration, and chemical mechanical polishing (CMP) of semiconductor materials. Additionally, she works with characterization techniques such as High-resolution X-Ray Diffraction (HRXRD) and Transmission Electron Microscopy (TEM).

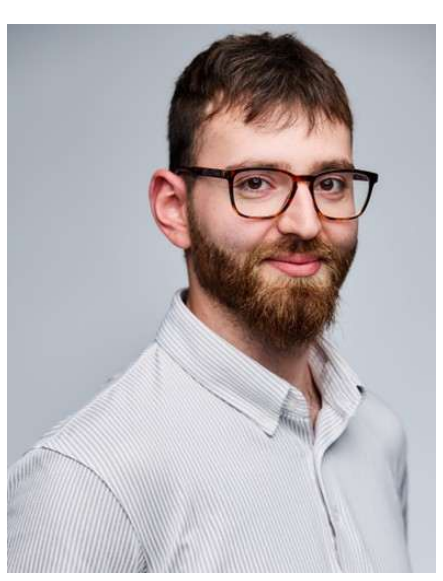

**Vakhtang Chulukhadze** (Student member, IEEE) received his B.S. from the University of Rochester and is currently pursuing his Ph.D. in electrical engineering at the University of Texas at Austin. His research has focused on lithium niobate microsystems spanning a wide range of frequencies and applications. His interests include further system integration of these piezoelectric components into larger systems for front-end signal processing and sensing. Vakhtang is the Jim and Dorothy Doyle scholar, as well as a Russel F. Brand Engineering Acoustics fellow, and Professor Edith Clarke fellow. He received an outstanding poster award at the Hilton Head workshop 2024 alongside the Ben Streetman award in graduate research on electronic and photonic devices.

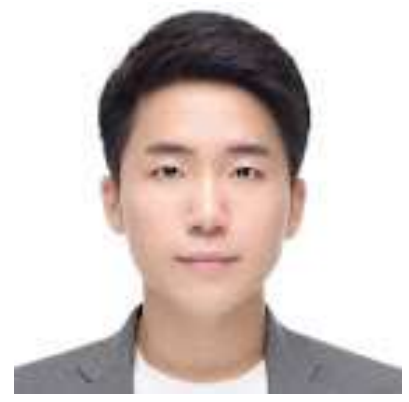

**Sinwoo Cho** (Student member, IEEE) is a Ph.D. candidate at UT Austin ECE. He received his B.S. degree in Biomedical Engineering from Inje University, Korea, in 2019 and his M.S. degree in Integrated Design Engineering from Keio University, Japan, in 2021. His research primarily focuses on RF acoustic front-end, including RF microsystem, RF MEMS, filter, and resonator. He received the Microwave Theory and Technology Society (MTT-S) Graduate Fellowship 2025, IEEE International Microwave Symposium third place Best Student Paper Award in 2024, IEEE Micro Electro Mechanical Systems (MEMS) Best Student Paper Finalists in 2021, and 2024 respectively, Qualcomm Innovation Fellowship (QIF) North America Finalists in 2022, and IEEE Texas Symposium on Wireless & Microwave Circuits and Systems (WMCS) Student Research Competition 1st Place Award in 2022.

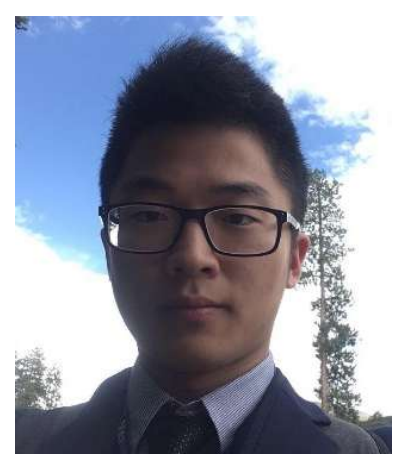

**Michael Evan Liao** is a research scientist at Apex Microdevices. Dr. Liao received his B.S. degree in Chemistry (2017), B.A. degree in Ethnomusicology Jazz Performance (2017), M.S. degree in Chemistry (2017), and Ph.D. in Materials Science & Engineering (2022) at the University of California Los Angeles. His research aims to develop commercial-compatible solutions towards fabricating composite wafers via wafer bonding of various semiconductor materials including, but not limited to, β-$Ga_2O_3$, SiC, and $LiNbO_3$. He received the IUCr Young Scientist Award in 2019 issued by the International Union for Crystallography at the ICCGE-19/OMVPE-19 conference and also the Best Student Paper Award in 2018 at the AiMES 2018 conference (Electrochemical Society). Dr. Liao joined Apex Microdevices after his tenure as a National Research Council Postdoctoral Fellow at the U.S. Naval Research Laboratory, Washington, DC.

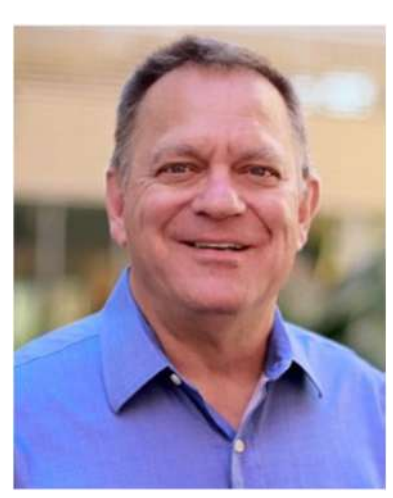

**Mark S. Goorsky** received his B.S. degree from Northwestern University in 1984 and his Ph.D. from the Massachusetts Institute of Technology in 1988. He has been on the faculty at UCLA since 1991 and was department chair from 2005-2010. His research focuses on wafer bonding, layer exfoliation and transfer, and chemical mechanical polishing of semiconductors and optical materials. Goorsky also provides expertise in materials characterization of semiconductor materials and devices, with emphasis on structural (x-ray scattering and electron microscopy) and chemical (electron energy loss spectroscopy, energy dispersive elemental analysis) techniques. His current research areas include materials integration of wide bandgap semiconductors and thin film electro-optic-acoustic materials such as lithium niobate. Goorsky received the university-wide 2016 UCLA Distinguished Teaching Award and the Harvey L. Eby Award for the “Art of Teaching”, was a member (2011-2015) of the US Air Force Science Advisory Board, was (2002-2019) associate editor for the Journal of Crystal Growth, was awarded the T.S. Walton Award from the Science Foundation of Ireland in 2010 and received (1995-2000) a National Science Foundation Career Award.

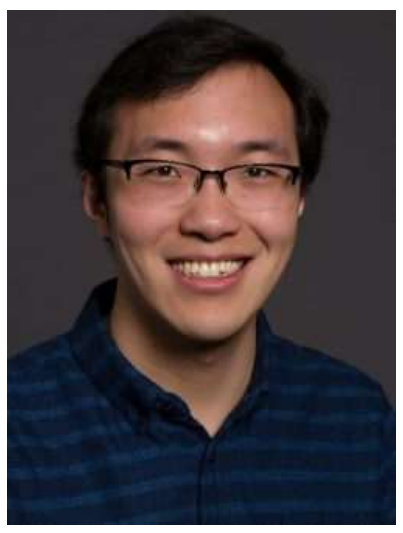

**Ruochen Lu** (Senior Member, IEEE) is an Assistant Professor in the Department of Electrical and Computer Engineering at The University of Texas at Austin. He received the B.E. degree with honors in microelectronics from Tsinghua University, Beijing, China, in 2014, and the M.S. and Ph.D. degree in electrical engineering from the University of Illinois at Urbana-Champaign, Urbana, IL, USA, in 2017 and 2019, respectively. His research primarily focuses on developing chip-scale acoustic and electromagnetic components and microsystems for RF applications. His works aim to demonstrate RF MEMS platforms, toward higher operating frequencies and more efficient transduction between the EM and acoustics. In addition, he works on ultrasonic transducers and multi-physics hybrid microsystems for signal processing, sensing, and computing applications. He received IEEE TC-S Microwave Award in 2022, IEEE Ultrasonics Early Career Investigator Award in 2024, and NSF Career Award in 2024. Along with his students, he received Best Student Paper Awards at IEEE International Frequency Control Symposium (IFCS) 2017 and 2023, IEEE International Ultrasonics Symposium (IUS) 2018 and 2022, IEEE International Conference on Microwave Acoustics & Mechanics (IC-MAM) 2022, and third place Best Student Paper Award at IEEE International Microwave Symposium (IMS) 2024. He received the Junior Faculty Excellence in Teaching Award in 2024 from the University of Texas at Austin. He is an associate editor of IEEE Journal of Microelectromechanical Systems, IEEE Transactions on Ultrasonics, Ferroelectrics, and Frequency Control, and a topic editor of IEEE Journal of Microwaves.